\documentclass[10pt,sigconf,letterpaper]{acmart}
\renewcommand\footnotetextcopyrightpermission[1]{} 
\begin{document}\sloppy

\title{Web Performance with Android's Battery-Saver Mode 
}

\author{Utkarsh Goel}
\affiliation{\institution{Akamai Technologies, Inc.}}
\email{ugoel@akamai.com}

\author{Stephen Ludin}
\affiliation{\institution{Akamai Technologies, Inc.}}
\email{sludin@akamai.com}

\author{Moritz Steiner}
\affiliation{\institution{Akamai Technologies, Inc.}}
\email{mosteine@akamai.com}

\pagenumbering{gobble}

\begin{abstract}
A Web browser utilizes a device's CPU to parse HTML, build a Document Object Model, a Cascading Style Sheets Object Model, and render trees, and parse, compile, and execute computationally-heavy JavaScript.
A powerful CPU is required to perform these tasks as quickly as possible and provide the user with a great experience. 
However, increased CPU performance comes with increased power consumption and reduced battery life on mobile devices.
As an option to extend battery life, Android offers a battery-saver mode that when activated, turns off the power-hungry and faster processor cores and turns on the battery-conserving and slower processor cores on the device. 
The transition from using faster processor cores to using slower processor cores throttles the CPU clock speed on the device, and therefore impacts the webpage load process.
\looseness -1

We utilize a large-scale data-set collected by a real user monitoring system of a major content delivery network to investigate the impact of Android's battery-saver mode on various mobile Web performance metrics. 
Our analysis suggests that users of select smartphones of Huawei and Sony experience a sudden or gradual degradation in Web performance when battery-saver mode is active.
Battery-saver mode on newer flagship smartphones, however, does not impact the mobile Web performance.
Finally, we encourage for new website design goals that treat slow~(and throttled-CPU) devices kindly in favor of improving end-user experience and suggest that Web performance measurements should be aware of user device battery charge levels to correctly associate Web performance.
\looseness -1
\end{abstract}
\maketitle

\section{Introduction}
\label{sec:introduction}
  
In the last several years, mobile websites have grown drastically in both complexity and size~\cite{tammy2,tammy1}. 
This growth has led to slower page loads and higher user frustration with the website~\cite{SmallScreensMightSales}. 
As websites continue to grow in complexity, the key to improve website responsiveness is to build faster networks, optimize website content, and produce mobile devices with faster CPUs. 
While many networks have already begun to deploy new infrastructure and support faster communication protocols~\cite{http2:rfc,Goelipv6,quic,nygren:ipv6}, and most websites already employ a suite of website optimization techniques~\cite{grigorik2013high}, the mobile device CPU remains a limiting factor to mobile Web performance~\cite{MoritzSmartphone15}. Unlike desktop and laptop CPUs, Mobile CPUs are designed for power efficiency.
\looseness -1

More specifically, Android smartphones have a battery-saver mode that lowers the battery consumption when the charge level drops below a certain threshold. 
Among other things, the battery-saver mode reduces the device's performance by throttling the CPU clock speed, where it deactivates the power-hungry and faster processor cores and activates the battery-saving and slower processor cores~\cite{cpuThrottle}. 
The process of loading Web pages includes HTML parsing, downloading and processing of JavaScript, Cascading Style Sheets~(CSS), image resources, executing computationally-heavy JavaScript, and building the Document Object Model (DOM), CSS Object Model~(CSSOM), and render tree~\cite{rendertree}.
Since all these tasks make use of the device's CPU resources, in this paper we seek to investigate whether or not a throttled CPU clock speed under an active battery-saver mode degrades the mobile Web performance for the end-user. 
Simply put, do mobile websites load slower on Android phones when the battery charge levels drop below a certain threshold?
\looseness -1

Since the impact of battery-saver mode on mobile Web performance has not garnered much developer interest in the past, our goal with this paper is to bring awareness to the developer community about potential performance impacts and thus motivate the need for new website design goals and decisions that treat mobile devices differently, especially when battery-saver modes are active and the CPU clock speeds are throttled.
We make the following contributions in the paper:
\looseness -1

\vspace{5pt}
\noindent
\textbf{Dataset Richness}: To investigate the impact of battery saver modes on mobile Web performance, we utilized a large-scale Web performance dataset collected by Akamai mPulse for websites loaded by real users on various mobile devices~\cite{mpulse}.
Our dataset contains various Web performance metrics collected for 10 million pages, loaded on 300 different smartphone models, connected to 81 cellular ISPs in 39 countries, from July 2017 to March 2018.

\vspace{5pt}
\noindent
\textbf{Inferences Drawn:} Using a large-scale mobile Web performance dataset, we discover that under battery-saver mode, select phones from Huawei, Sony Xperia, and Samsung Galaxy series degrade mobile Web performance metrics, such as the page load time~(PLT), time to first paint~(TTFP), total LongTask time, time to interactive~(TTI), and frame rate~\cite{nt,continuity,lt,fr}. 
The data also suggest that the battery-saver mode makes a higher impact on Web performance when Web pages load in faster mobile network conditions. 
The above findings demonstrate a clear need for new website design goals that would go hand-in-hand with the understanding of how user devices with low battery charge levels could degrade the mobile Web experience. 
Specifically, developers might want to build websites that adapt to different battery situations to overcome any user-perceived responsiveness issues inflicted by the battery-saver mode.
Additionally, to reduce utilization of the throttled CPU, users may use browsers that offload CPU-intensive computations to the cloud~\cite{amazonsilk,operamini,puffin}.
\looseness -1

The rest of the paper is organized as follows.
Section~\ref{sec:background} gives a background on Android's battery-saver modes.
In Sections~\ref{sec:data} and~\ref{sec:perf}, we discuss our data collection methodology and the impact of battery-saver mode on various Web performance metrics. 
In Section~\ref{sec:related_work}, we discuss the related work.
Finally, we conclude in Section~\ref{sec:conclusions}.
\looseness -1

\section{Background}
\label{sec:background}
 
The battery-saver mode, when activated, reduces the screen brightness, limits the use of WiFi, disables power-consuming location-sharing services, and reduces the application background activity. 
Android smartphones, such as LG G5, Huawei Y6 Elite, Alcatel Pixi 4, and many others, provide two user-configurable options for the battery charge level threshold at which the battery-saver mode turns on automatically~\cite{y6elite,alcatel,lgg5,android}. 
These threshold values are 5\% and 15\%. 
However, other Android smartphones, such LG G3 and LG VS985, provide four user-configurable options for this threshold, which are 10\%, 20\%, 30\%, and 50\%~\cite{g3}. 
Similarly, Samsung phones, such as Galaxy S6, Galaxy J5, and others, also provide four user-configurable options for this threshold, which are 5\%, 10\%, 20\%, and 50\%~\cite{j5,s6}. 
Note that even though there is an automated way to turn on the battery-saver mode every time the battery charge level drops below a certain threshold, the activation of this feature on the device depends on a user's interest in saving battery charge. 
\looseness -1

Other smartphones, such as Samsung Galaxy S8, Note 8, and Galaxy S5, do not provide a way to set a threshold, which means that the activation and deactivation of the power-saving modes must be done manually by users every time they want to save power~\cite{s5,s8}. 
The activation of the power-saving modes on these devices may even be lower compared to the devices that offer a threshold for automatically activating the power-saving mode.
\looseness -1

Sony's Xperia Z5 Compact has several power-saving modes, such as the Doze mode, that turns on automatically to save power when the device screen is turned off~\cite{z5}. 
Other power-saving modes, such as the Stamina mode, allow users to set a battery charge level threshold at which power-saving mode activates. 
Moreover, the power-saving mode on Sony Xperia Z5 Compact also allows users to decide whether or not the CPU clock speed should be throttled, in addition to, or instead of, disabling mobile data and WiFi~\cite{z5_mode}. 
As such, some users may choose to activate power-saving mode without throttling the CPU clock speed, while others may choose to throttle the CPU clock speed.
\looseness -1

\section{Data Collection Methodology}
\label{sec:data}

To analyze the different performance metrics pertaining to websites loaded on various real user smartphones, we utilized the data collected by Akamai's mPulse product~\cite{mpulse}. 
mPulse embeds a lightweight JavaScript snippet to some HTML responses and leverages the Web browser-exposed Navigation Timing API to collect performance-related information to estimate the time taken to load a page~\cite{nt}. 
mPulse also utilizes the browser exposed Battery Status API to associate the measured website performance data with the device's battery charge level information at the time of measurement~\cite{bsa}.
The collected data is reported back to Akamai servers~\cite{akamai}, which we analyze to assess the impact of battery saver modes on page load performance.
Our data consists of a total of 10 million page load transactions for 480 unique websites loaded on 300 different smartphone models connected to 81 cellular ISPs in 39 countries from July 2017 through March 2018.
\looseness -1

To investigate the Web performance experienced on devices with low battery charge levels, we calculate the median page load time~(PLT) observed for page loads pertaining to 6-field buckets, comprised of: 1)~the country name in which the page was loaded, 2)~the ISP over which the page was loaded~(note that the first two pieces of information are deduced from the MaxMind's database of mapping IP addresses to geographical locations~\cite{maxmind}), 3)~the URL of the website loaded, 4)~the smartphone model used to load the website, 5)~the HTTP protocol version (HTTP/1.1 or HTTP/2) used, and 6)~the device battery percentage, ranging from 1 to 100\%.
Note that we refer to PLT as the time from the start of page navigation until the \texttt{loadEventStart} event is triggered by the Web browser~\cite{nt}.
Also note that bucketing the data with this approach helps to precisely understand how fast a given website loads on a given ISP network under specific constraints, such as the user's country, user's ISP, website URL, the smartphone, and the HTTP protocol, and thus mitigate the influence of many factors that might affect the analysis of Web performance. 
\looseness -1

In addition to calculating the median PLT for beacons in each bucket, we calculate the 10\textsuperscript{th}, 25\textsuperscript{th}, 75\textsuperscript{th}, and 90\textsuperscript{th} percentile PLT values. 
By calculating these percentile values for each of the 6-field bucket, we could assume that the 10\textsuperscript{th} percentile PLT tends to represent page loads in fast cellular networks, and that the 90\textsuperscript{th} percentile PLT tends to represent page loads in slower cellular networks. 
This categorization would help us understand how the impact of battery-saver mode changes as network performance improves or degrades.
\looseness -1

Finally, similarly to how we calculate PLT distributions, we also calculate the TTFP observed for loading different websites on different smartphones. 
Additionally, we collect the total LongTask time, the time to interactive, and the frame rate observed when loading websites under various battery charge levels. 
Note that since the LongTask, time to interactive, and frame rate metrics depend on the device hardware, as opposed to the network performance, for these metrics we do not calculate 10\textsuperscript{th}, 25\textsuperscript{th}, 50\textsuperscript{th}, 75th, and 90\textsuperscript{th} percentile values~(because we represent percentile values as network speed) but instead plot the whole distributions in appropriate figures.
\looseness -1

Note that the dataset we prepared consists of 25 unique combinations~(comprising of country, ISP, URL, smartphone name, and HTTP protocol) for which mPulse library was executed on at least 100 page loads for each of the battery charge levels ranging from 1\% to 100\% - a total of at least 10,000 beacons for each combination. 
Additionally, the dataset consists of 63 unique combinations for which mPulse library was executed on at least 100 page-load transactions, for at least 90 battery charge levels  - a total of at least 9,000 beacons for each combination. 
The low number of combinations observed in the dataset is a consequence of the fact that only about 2.5\% of the total page loads occurred when battery charge levels were below 15\%, which yielded less than 100 beacons for some battery charge levels on some combinations.
Additionally, the low number of page loads under battery charge levels less than 15\% may suggest that either most users keep their phones charged over 15\% at all times or that when battery charge levels drop below 15\%, users reduce their Web browsing activities in favor of conserving battery charge.
\looseness -1

\section{Measuring Website Performance}
\label{sec:perf}
 
To estimate the performance of a website, we analyze several Web performance metrics under different device battery charge levels, such as PLT and TTFP -- the time since the start of the page navigation until the browser paints the first pixels. 
We also analyze the total LongTask time -- the time during which the browser main/UI thread is blocked and therefore, the user cannot interact with the page~\cite{chromeThreads}. 
Finally, we measure the average rate of printing frames on the screen~\cite{fr}.  
\looseness -1

\subsection{Measuring Page Load Time~(PLT)}
 
\subsubsection{Performance on Huawei Y6 Elite}
\begin{figure}[t]
 \centering
 \minipage{0.47\textwidth}
\includegraphics[width=\linewidth]{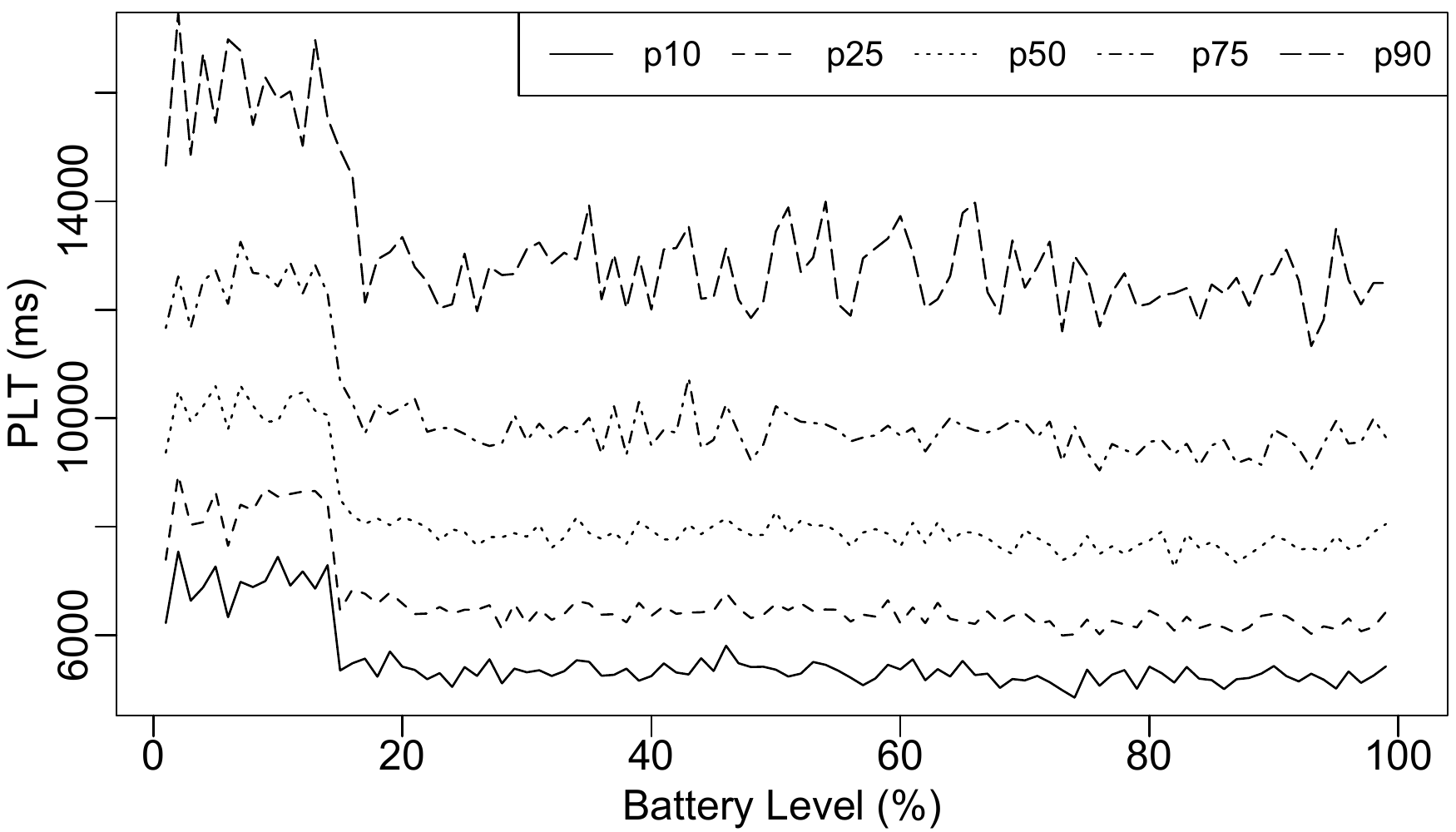}
 \vspace{-15pt}
 \captionsetup{width=1\textwidth}
\caption{PLT distributions across different device battery charge levels, as measured for a page loaded on Huawei Y6 Elite mobile device.}
 \label{fig:www_vodafone_com_au_y6Elite}
 \endminipage
\end{figure}

In Figure~\ref{fig:www_vodafone_com_au_y6Elite}, we show various percentile PLT values for a page loaded on Huawei Y6 Elite smartphone. 
From the figure we can observe that across all percentiles, the PLT inflates as soon as the battery charge level drops to 15\%, likely due to degraded CPU clock speeds triggered by the battery-saver mode. 
The figure also suggests that, on most Huawei Y6 Elite smartphones, the default threshold for when the battery-saver mode initiates is set to 15\%.
\looseness -1

In the Table~\ref{tbl:one}, we summarize the analysis from Figure~\ref{fig:www_vodafone_com_au_y6Elite} to compare the PLT observed when battery charge levels are, for example, 8\% and 50\%. Note that the battery charge levels 8\% and 50\% are just two example representative points for comparing the performance under the battery-saver and normal modes.
\looseness -1

\begin{table}[]
 \centering
 \minipage{0.47\textwidth}
\includegraphics[width=\linewidth]{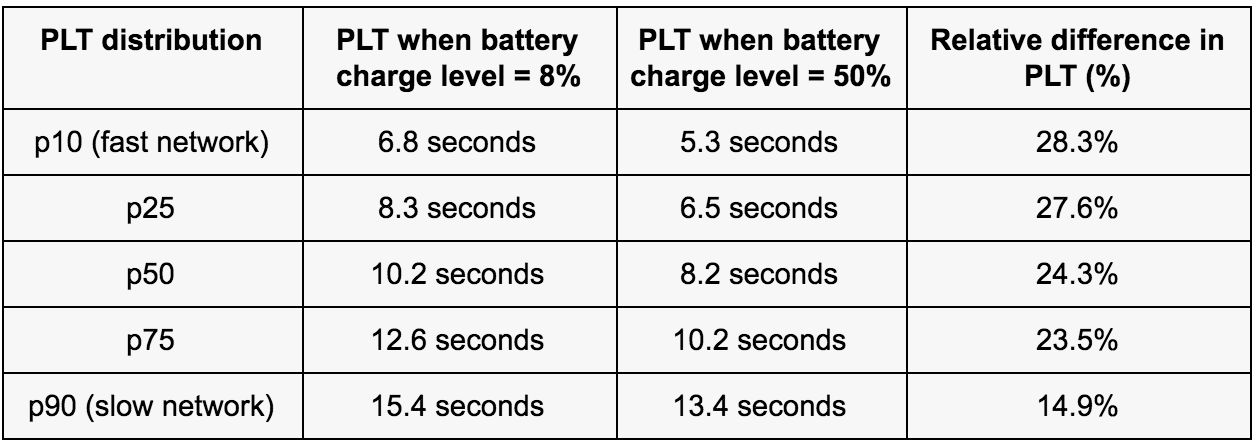}
 \vspace{-5pt}
 \captionsetup{width=1\textwidth}
\caption{Summary of PLT distributions on Huawei Y6 Elite phone. 
}
 \label{tbl:one}
 \endminipage
\end{table}

From the table we can observe that the relative difference in the page load decreases as the network performance degrades. 
Specifically, among the two battery levels, 8\% and 50\%, for page loads represented by the 10\textsuperscript{th} percentile distribution (p10), the PLTs at 8\% are higher by 28\% compared to the PLTs at 50\%. 
Whereas, for the 90\textsuperscript{th} percentile distribution (p90), the PLTs at 8\% are higher by 15\% compared to the PLTs at 50\%. 
This downward trend suggests that when websites are loaded on faster networks, the smartphone's CPU becomes a more significant bottleneck for website performance. 
Indeed this trend is similar to a previous research study that investigates how device performance impacts Web performance~\cite{MoritzSmartphone15}.
\looseness -1

Both PLT and TTFP metrics make direct impact on the user experience. As such, an increase in these metrics, due to degraded CPU clock speeds, represents a degradation in the user experience.
\looseness -1

\begin{figure}[t]
 \centering
 \minipage{0.47\textwidth}
\includegraphics[width=\linewidth]{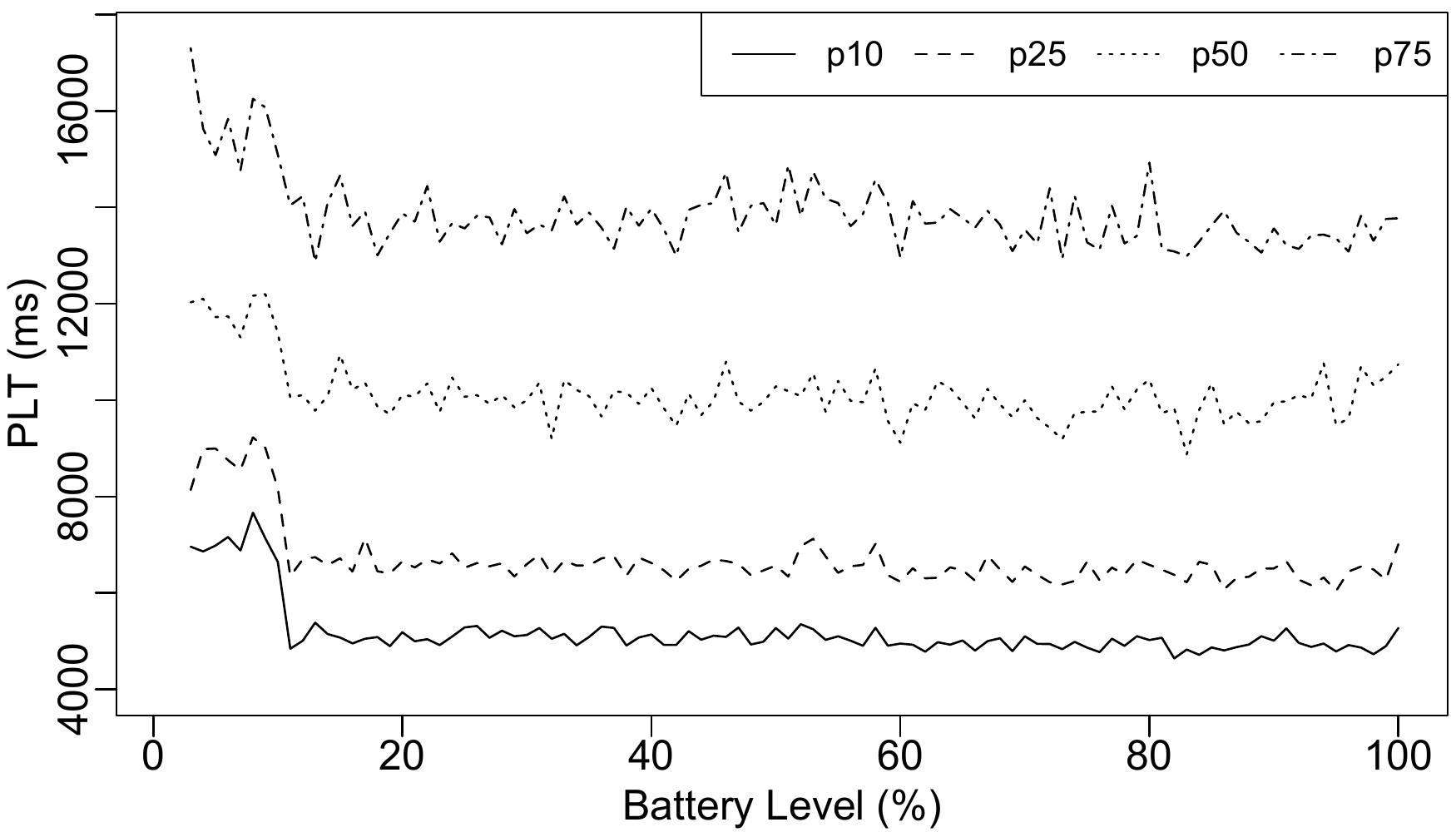}
 \vspace{-15pt}
 \captionsetup{width=1\textwidth}
\caption{PLT distributions across different device battery charge levels, as measured for a page loaded on Huawei P8 Lite (2015) mobile device.}
 \label{fig:vodafone_ro_content_P8Lite_h1}
 \endminipage
 \vspace{-10pt}
\end{figure}

\subsubsection{Performance on Huawei P8 Lite (2015 model)}
Similarly to Figure~\ref{fig:www_vodafone_com_au_y6Elite}, as shown in Figure~\ref{fig:vodafone_ro_content_P8Lite_h1}, we observe a sudden inflation in PLTs when loading a Web page on the 2015 model of Huawei P8 Lite smartphone. Since the inflation occurs when the battery charge level is at 10\%, it appears that the battery-saver mode on the Huawei P8 Lite smartphone turns on for most users at 10\%.
\looseness -1

\begin{figure}[]
 \centering
 \minipage{0.47\textwidth}
\includegraphics[width=\linewidth]{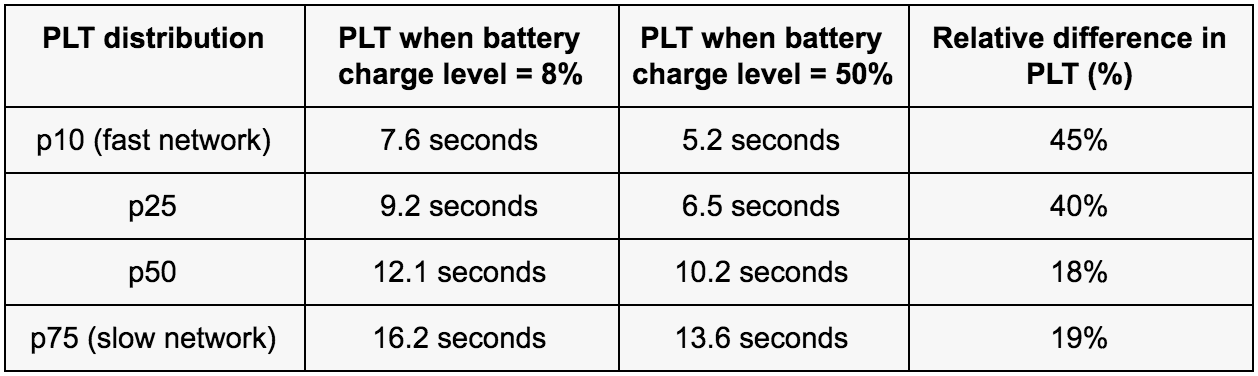}
 \vspace{-10pt}
 \captionsetup{width=0.9\textwidth}
\caption{Summary of PLT distributions on Huawei P8 Lite phone. 
}
 \label{tbl:two}
 \endminipage
 \vspace{-10pt}
\end{figure}

Additionally, as shown in Table~\ref{tbl:two}, when we compare the relative PLT differences across different distributions, we can observe that the impact of throttled CPU clock speeds on PLT appears to be more significant when pages are loaded in fast network conditions.
\looseness -1

We observed similar trends in performance for other websites that loaded on Huawei P8 Lite smartphone, however, we do not show them due to page limits. 
Additionally, note that for websites loaded on the 2017 model of Huawei P8 Lite smartphone, we did not observe sudden inflation in PLT or TTFP metrics at any battery charge level, potentially because the 2017 model has a faster CPU that does not impact the page load despite the drop in CPU performance~\cite{huaweicomparison}. 
Finally, we noticed that newer smartphones from the same family, such as Huawei P9 Lite and P10 Lite, also do not degrade website performance under low battery charge levels.
\looseness -1

\begin{figure}[t]
 \centering
 \minipage{0.47\textwidth}
\includegraphics[width=\linewidth]{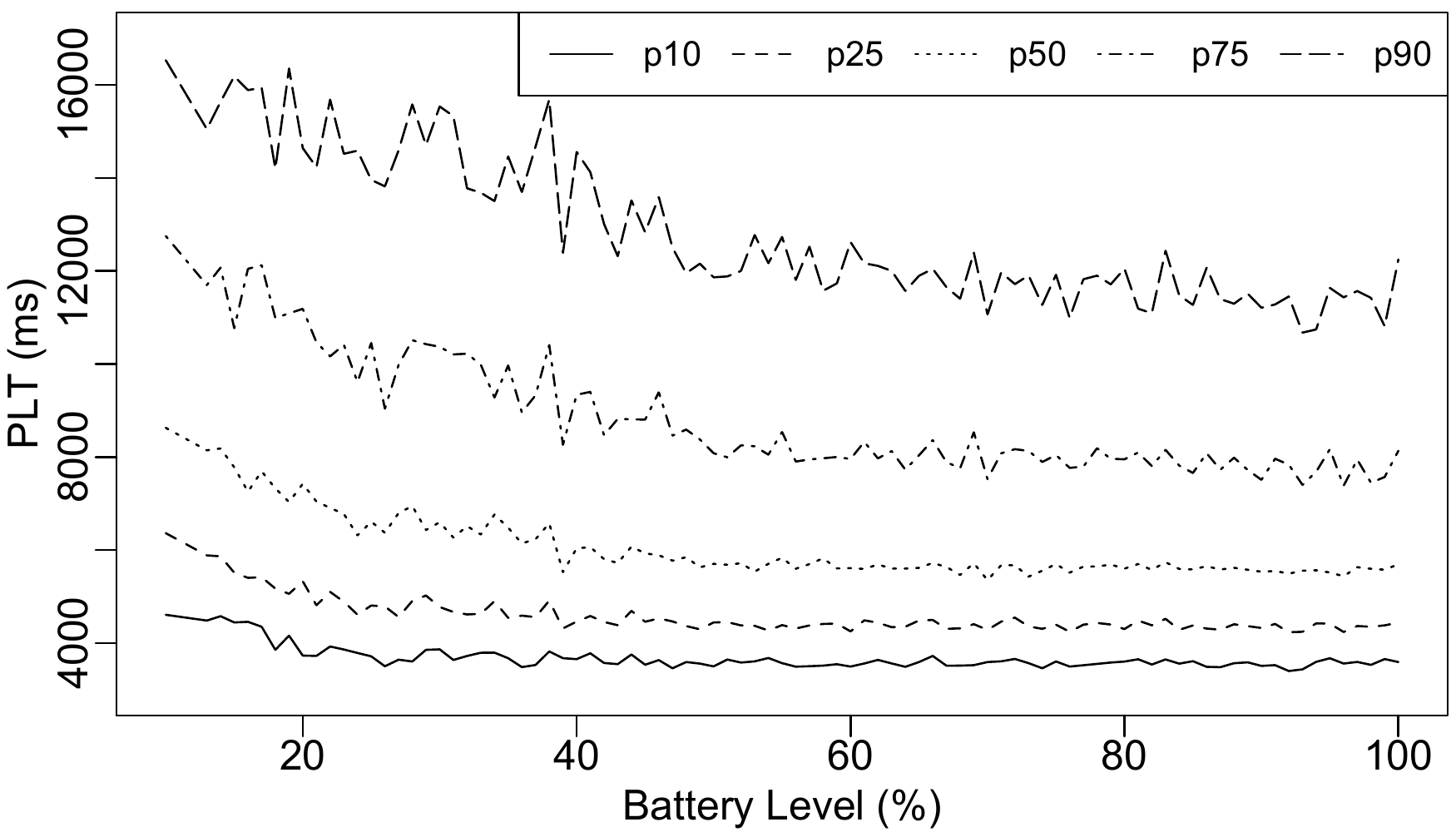}
 \vspace{-15pt}
 \captionsetup{width=1\textwidth}
\caption{Page load time distributions across different device battery charge levels, as measured for a page loaded on Sony Xperia Z5 Compact mobile device.}
 \label{fig:jp_rakuten_XperiaZ25_h2_ntt_8m}
 \endminipage
 \vspace{-10pt}
\end{figure}

\subsubsection{Performance on Sony Xperia Z5 Compact}

In Figure~\ref{fig:jp_rakuten_XperiaZ25_h2_ntt_8m} we show that when pages load on Sony Xperia Z5 Compact smartphone, there exists an upward slope indicating that page load times continue to rise as battery charge levels drop below 40\%.
We have not been able to identify the cause for such an upward slope, as opposed to a sudden increased in page load time, even though this device allows users to set a threshold at which the battery-saver mode activates. Perhaps, this device has a linear slowdown in the CPU clock speed when the battery charge levels reach 40\%. 
Note that we did not observe such a strong upward slope for other Sony devices, such as Xperia X Compact, Xperia X Performance, and Xperia XZ and in fact, there was no inflation in PLT for these devices under low battery charge levels.
\looseness -1

\subsubsection{Performance on High-end Phones}

\begin{figure}[t]
 \centering
 \minipage{0.47\textwidth}
\includegraphics[width=\linewidth]{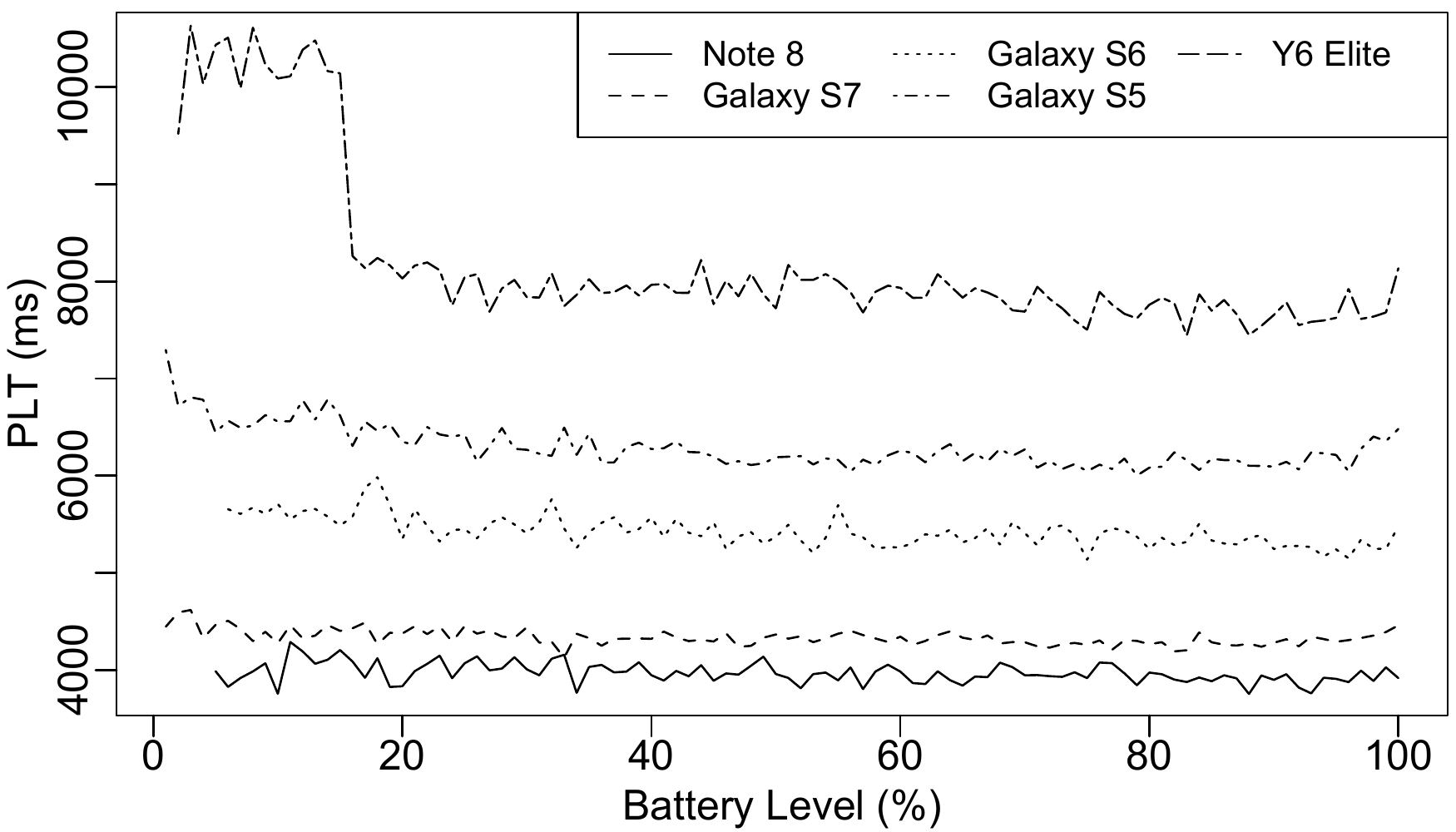}
 \vspace{-15pt}
 \captionsetup{width=1\textwidth}
\caption{PLT distributions across different device battery charge levels, as measured for a page loaded on various Samsung smartphones and Huawei Y6 Elite.}
 \label{fig:www_vodafone_com_au_several_devices_median}
 \endminipage
 \vspace{-5pt}
\end{figure}

We compare the website performance for pages loaded on various new and old Samsung flagship smartphones under different battery charge levels. 
Note that for the purposes of comparing performance across different Samsung devices, in Figure~\ref{fig:www_vodafone_com_au_several_devices_median} we only show the PLT values calculated for the 50\textsuperscript{th} percentile distribution at various battery charge levels. 
Additionally, note that for making the comparison easier to understand, in Figure~\ref{fig:www_vodafone_com_au_several_devices_median}, we compare the performance observed for Huawei Y6 Elite smartphone for the same website as the one used in Figure~\ref{fig:www_vodafone_com_au_y6Elite}.
\looseness -1

From Figure~\ref{fig:www_vodafone_com_au_several_devices_median}, we observe that when the website is loaded on newer devices with high-performance CPU, the PLT gets lower across all battery charge levels.
For example, the PLTs on Note 8 are about 4 seconds, whereas the PLTs on Galaxy S6 are about 5.5 seconds. 
Similarly, PLTs on Galaxy S5 are about 6.5 seconds, and the PLTs on Y6 Elite under non-throttled CPU hardware are about 8 seconds. 
The performance differences across devices show that the website performance can be improved by upgrading the device hardware, regardless of the battery charge level.
\looseness -1

\begin{figure}[t]
 \centering
 \minipage{0.47\textwidth}
\includegraphics[width=\linewidth]{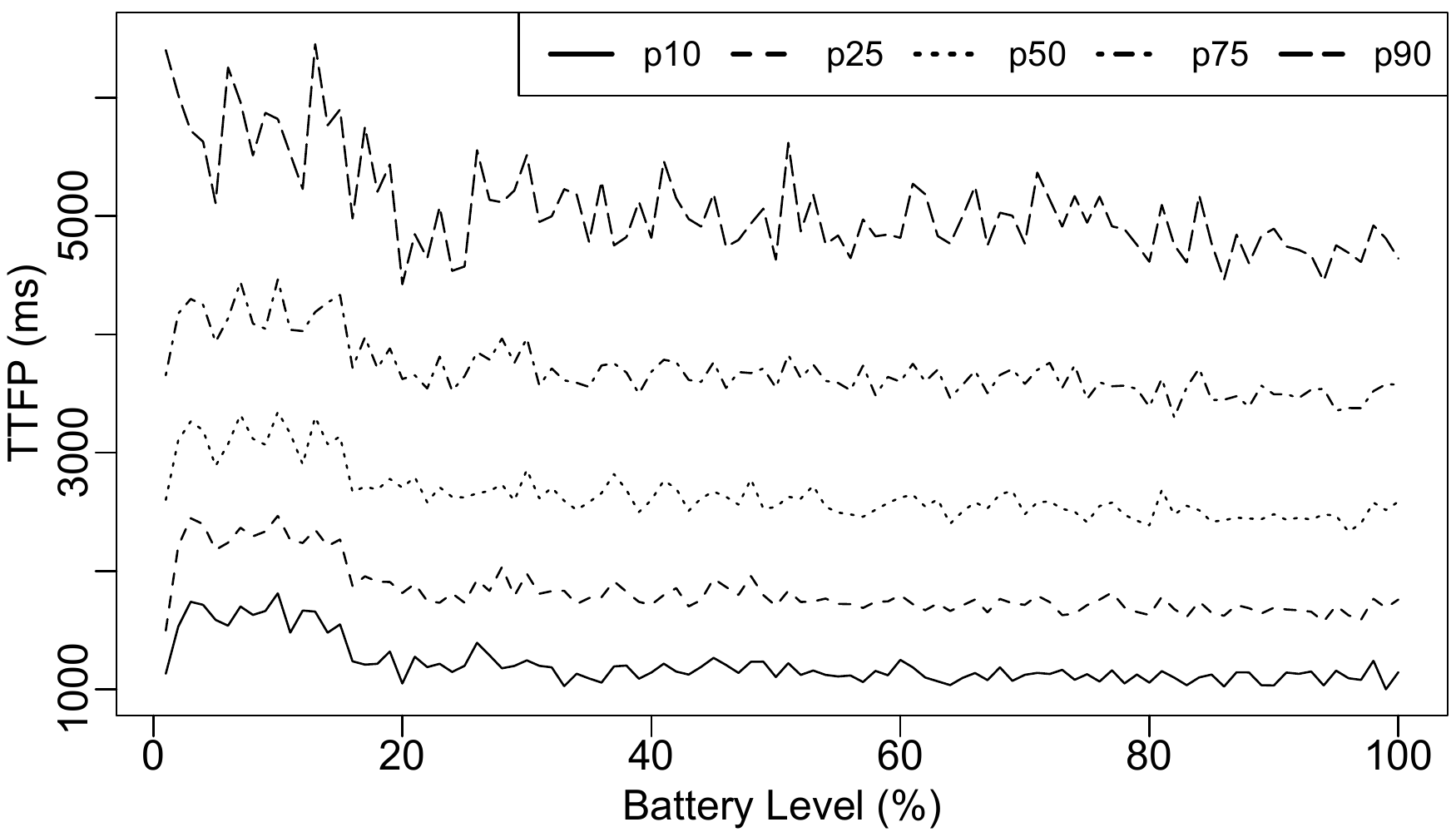}
 \vspace{-15pt}
 \captionsetup{width=1\textwidth}
\caption{TTFP distributions across different device battery charge levels, as measured for a page loaded on Huawei Y6 Elite mobile device.}
 \label{fig:www_vodafone_com_au_y6Elite_ttfp}
 \endminipage
 \vspace{-10pt}
\end{figure}

Additionally, from Figure~\ref{fig:www_vodafone_com_au_several_devices_median} we observe that PLTs on Note~8 do not experience sudden or gradual increase at any battery charge level.
Since Samsung Note 8 does not provide a default user configurable option to enable the battery-saver mode when the battery charge level reaches a certain threshold and that users must manually activate the battery-saver mode, perhaps most users tend to not activate the battery-saver mode and therefore we do not observe any inflation in PLTs.
Alternatively, since Galaxy Note 8 has octa-core processors built-in, perhaps even when the battery-saver mode is active, the throttled CPU clock speed is high enough to not negatively impact the page load process or the PLT~\cite{note8octacore}. 
However, we acknowledge that a better understanding of the inner workings of the device would help bear this out. 
For other devices, such as Galaxy S7 and S6, we do not observe any inflation in PLT regardless of the battery charge level. 
Similarly to Note 8, perhaps the throttled CPU clock speeds on these devices are high enough to not impact the PLT. 
\looseness -1

\subsection{Measuring TTFP}
 
In Figure~\ref{fig:www_vodafone_com_au_y6Elite_ttfp}, we show various percentile TTFP values for the same page as Figure~\ref{fig:www_vodafone_com_au_y6Elite}, when loaded on Huawei Y6 Elite smartphone.
From the figure we can observe that there exists a sudden inflation in the TTFP when the battery charge levels drop below 15\%. 
This trend suggests that the time when the browser paints the first pixels also increases when the device enables the battery-saver mode. 
Similarly to Figure~\ref{fig:www_vodafone_com_au_y6Elite}, we also noticed sudden inflation in TTFP at battery charge level 15\%, when loading websites on the 2015 model of Huawei P8 Lite smartphone.
\looseness -1

\begin{figure}[t]
 \centering
 \minipage{0.47\textwidth}
\includegraphics[width=\linewidth]{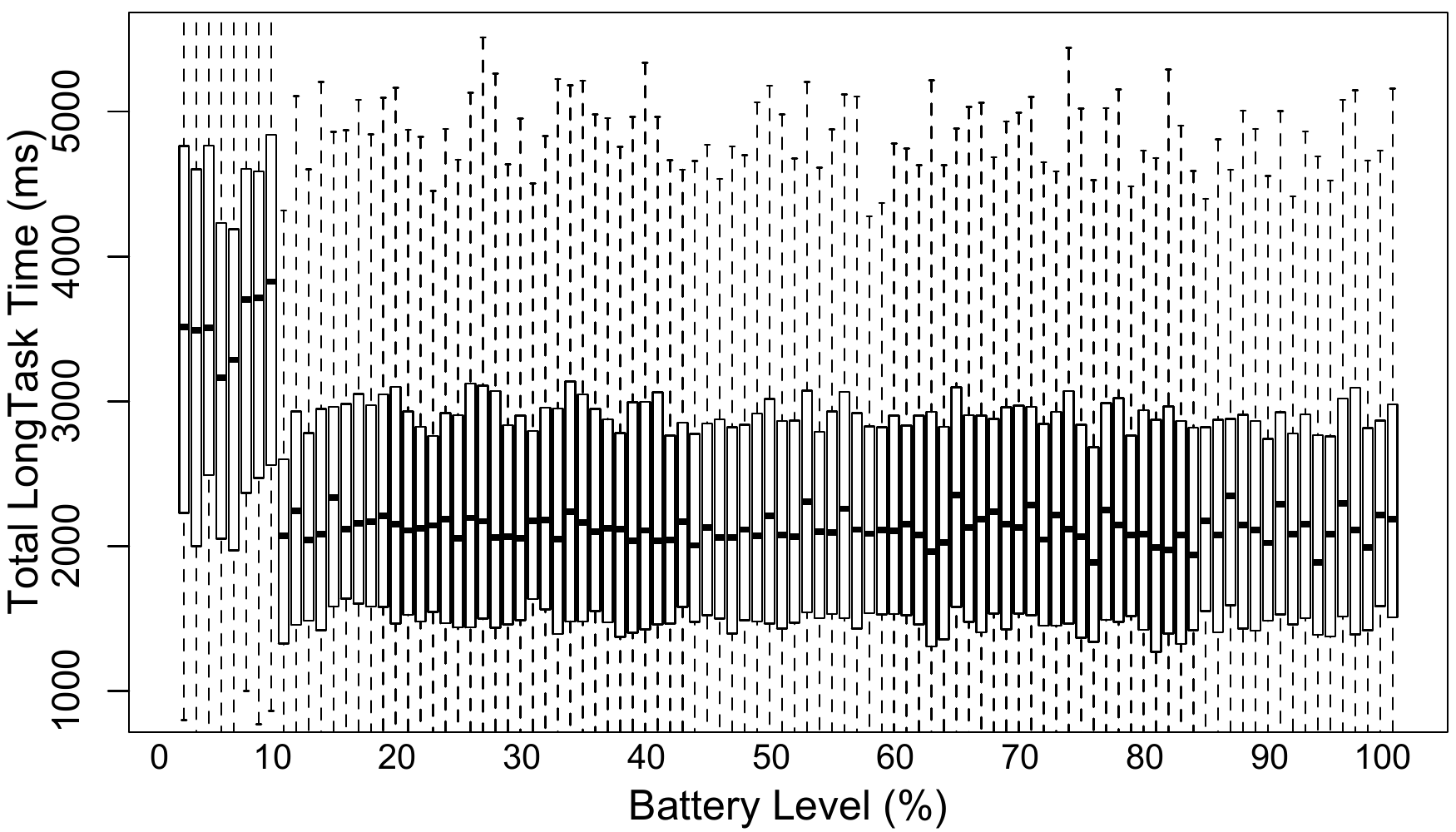}
 \vspace{-15pt}
 \captionsetup{width=1\textwidth}
\caption{LongTask time distributions across different device battery charge levels, as measured for a page loaded on P8 Lite (2015) smartphone. 
}
 \label{fig:www_bol_com_p8lite_boxplot}
 \endminipage
 \vspace{-10pt}
\end{figure}

\subsection{Measuring LongTask Time}
 
LongTask is a relatively new Web performance measurement API that allows identification of resources that make websites unresponsive to user interactions~\cite{lt}. 
More specifically, Web developers could use the LongTask API to detect the presence of tasks that block the browser UI/main thread for at least 50 milliseconds. 
When a website loads resources that block the browser UI/main thread, the user is unable to interact with the page. 
Specifically, a long task prevents the page from responding to user actions, such as scroll, click, tap, key, wheel, etc, until the long task has finished executing. 
This is because when a long task is executing, all user actions are queued behind the long task. 
\looseness -1

\begin{figure}[t]
 \centering
 \minipage{0.47\textwidth}
\includegraphics[width=\linewidth]{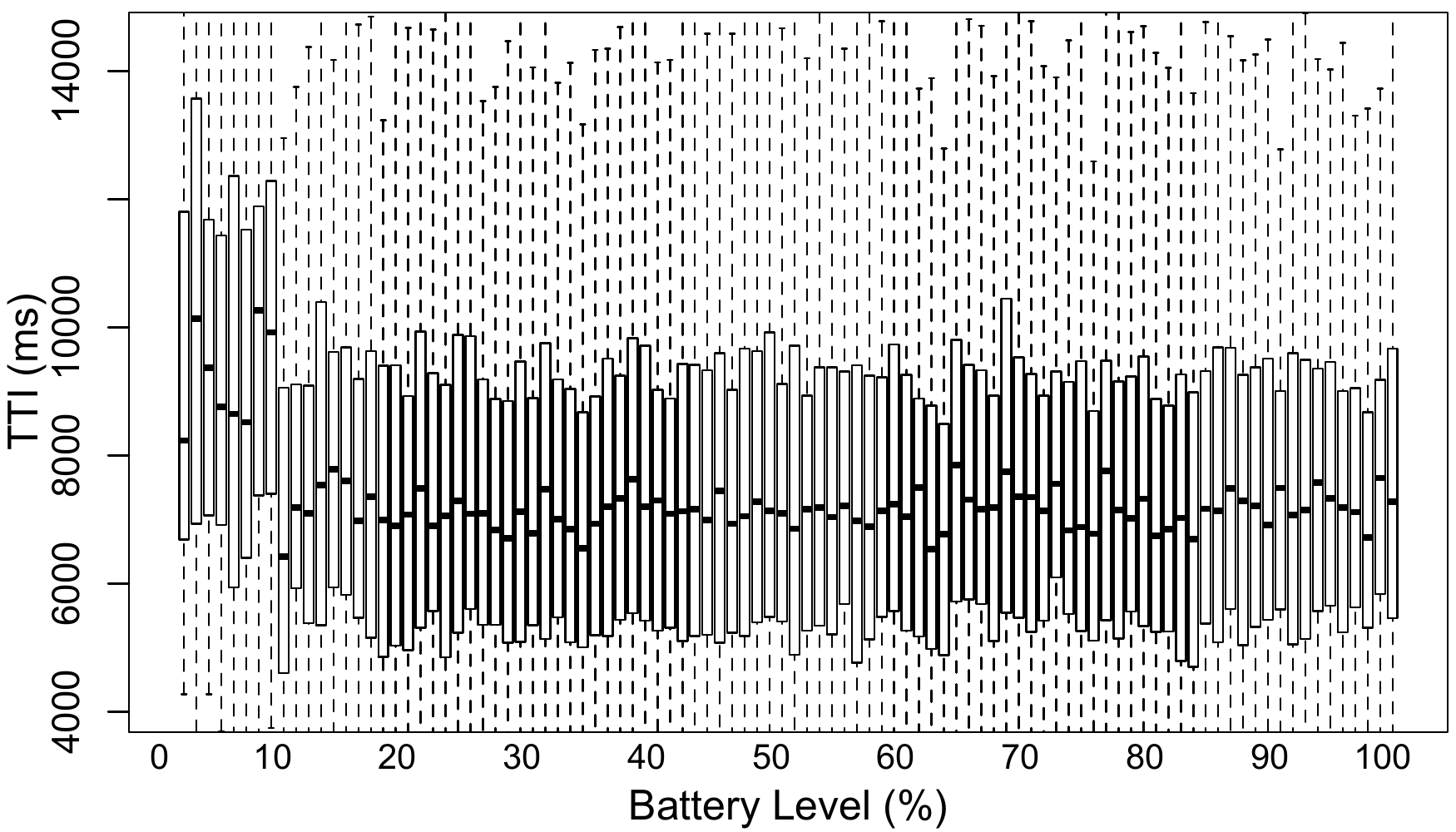}
 \vspace{-15pt}
 \captionsetup{width=1\textwidth}
\caption{TTI distributions across different device battery charge levels, as measured for a page loaded on P8 Lite (2015) smartphone.}
 \label{fig:www_bol_com_p8lite_boxplot_tti}
 \endminipage
 \vspace{-10pt}
\end{figure}

\begin{figure}[t]
 \centering
 \minipage{0.47\textwidth}
\includegraphics[width=\linewidth]{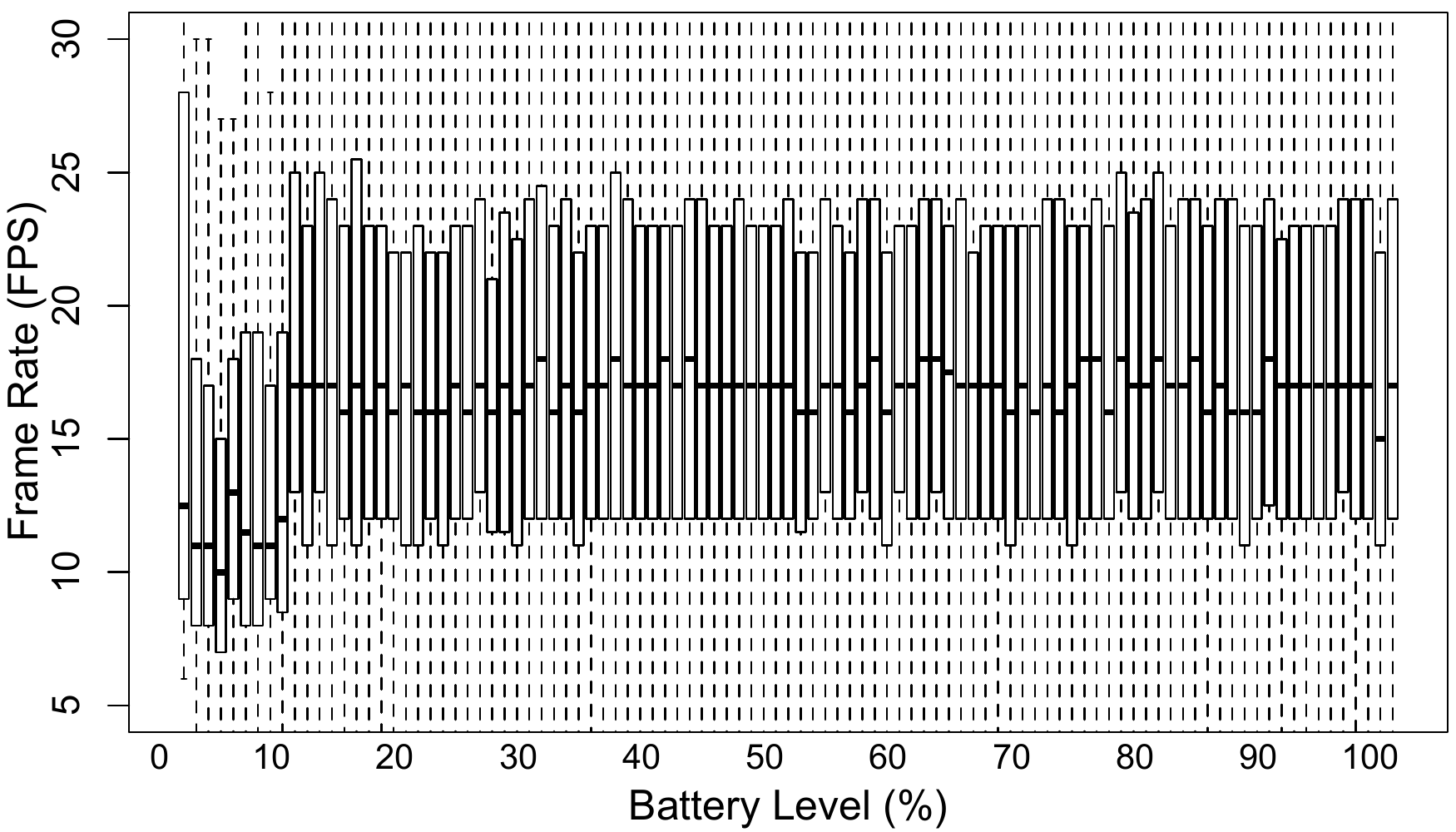}
 \vspace{-15pt}
 \captionsetup{width=1\textwidth}
\caption{FPS distributions across different device battery charge levels, as measured for a page loaded on P8 Lite (2015) smartphone.}
 \label{fig:www_bol_com_P8_Lite_boxplot_fr}
 \endminipage
 \vspace{-10pt}
\end{figure}

Poorly designed JavaScript code is one example of what might cause a browser main thread to block for over 50ms~\cite{poorJs}. 
Since the current LongTask API (v1) does not reveal the URL of the long task~(though the second version of the API will provide such detail, including the line number that caused the long task~\cite{ltv2}), it is unclear as to what particular resources block the browser main thread. 
Therefore, we only focus on investigating whether or not we observe a rise in the total LongTask time when device CPU clock speed reduces when battery charge levels drop below a certain threshold, as opposed to discussing the root cause of a large or small total LongTask time.
Also note that since LongTask time, time to interactive, and the frame rate metrics do not depend on network speed and instead depend on the device CPU performance, we use boxplot distributions in Figures~\ref{fig:www_bol_com_p8lite_boxplot}-\ref{fig:www_bol_com_P8_Lite_boxplot_fr} instead of plotting different percentile values as we did for previous graphs.
\looseness -1

In Figure~\ref{fig:www_bol_com_p8lite_boxplot}, we show the boxplot distributions of the total LongTask time across different battery charge levels when loading a website on the 2015 model of the Huawei P8 Lite smartphone.
From the figure we can observe that the total LongTask time inflates when device battery charge level drops below 10\%. 
The rise in the total LongTask time indicates that when the device CPU clock speed is throttled to minimize battery consumption, LongTasks block the main thread for longer than usual time. 
Note that we did not observe inflation in the total LongTask time on the 2017 model of Huawei P8 Lite smartphone, likely due to the fact that faster processors on the device do not impact the total LongTask time when their speeds are throttled by the battery-saver mode.
\looseness -1

\subsection{Measuring Time to Interactive~(TTI)}
 
Not only does a LongTask delay user interactions, but events callbacks (such as onLoad) are also delayed. 
In many pages where many LongTask exists as the page loads, the time at which the user could first interact with the page could also get delayed.
Additionally, note that even though LongTask is the prime cause for poor responsiveness to user interactions, other tasks, such as image decoding, heavy rasterization work, or presence of many layers on the page can also cause poor responsiveness~\cite{raster}.
\looseness -1

To investigate whether there is any impact of battery-saver mode on the time when the user could first interact with the website, we take a look at the time to interactive (TTI) metric.
The TTI metric is calculated based on when the page was visually ready for the user and when the page was ready for interaction~\cite{continuity}. 
Specifically, the former is calculated by calculating the maximum of time to first paint and time to \texttt{domContentLoadedEventEnd} event~\cite{dcl}. 
Once the time to visually ready is calculated, the first time period of 500 milliseconds during which the browser UI/main thread was idle marks the TTI for the page.
\looseness -1

In Figure~\ref{fig:www_bol_com_p8lite_boxplot_tti}, we show the boxplot distributions for TTI values observed for loading a website on Huawei P8 Lite smartphone under different battery charge levels. 
From the figure we can observe that the TTI values inflate as soon as the battery charge levels drop below 10\%. 
The sudden rise in TTI values indicates that users on this smartphone, and other similar smartphones, likely experience janks when interacting with the website, thus leading to poor user experience.
\looseness -1

\subsection{Measuring Frame Rate per Second}

The new \texttt{requestAnimationFrame} API allows Web developers to strategically schedule paint events on the screen with the goal of achieving a high frame rate during the website load~\cite{raf}. 
Typically, 60 frames per second (FPS) is considered ideal for good user experience, which means that the browser has exactly 16.6~ms to produce a frame. 
However, if the browser needs to perform tasks that delay the frame generation, the FPS declines. 
Note that a low frame rate can degrade the user experience, because under low FPS the page becomes unresponsive to user interactions. 
Therefore, we investigate whether or not the battery-saver mode impacts the frame rate observed across different page loads. 
Using the mPulse Continuity plugin we gathered the frame rate observed under various battery charge levels~\cite{continuity}. 
Note that we show analysis of frame rates only for one of the devices for which we observe a sudden rise in both the PLT and TTFP.
\looseness -1

As shown in Figure~\ref{fig:www_bol_com_P8_Lite_boxplot_fr}, the FPS observed on the 2015 model of Huawei P8 Lite drops as soon as the device battery charge levels drop below 10\%.
This indicates that for websites loading on this device, as the CPU performance degrades, the rate at which the browser paints on the screen also declines - leading to a potentially poor user experience.
\looseness -1

\section{Related Work}
\label{sec:related_work}
Several tools and studies have relied on active measurements to identify Web performance bottlenecks~\cite{gomez,catchpoint,wpt,wprof}.
Unlike passive measurements via real user monitoring systems~\cite{dynatrace,mpulse}, active measurement practices inherit several limitations because pages loaded in controlled environments do not represent the characteristics of how pages load in the real world~\cite{meenan}.
Steiner~\textit{et\,al.} used a real user monitoring system to compare the impact of CPU processors embedded in old and new smartphones on the web performance -- suggesting that page loads on new devices with fast CPUs are significantly faster than page loads on old devices with slow CPU~\cite{MoritzSmartphone15}.
Like other studies~\cite{mobilyzer}, the authors observed that faster CPUs on newer smartphones load webpages faster than those on old phones.
Another study reveals that about 35\% of the PLT is spent performing CPU-intensive tasks on user devices~\cite{wprof}.

\section{Discussion and Conclusion}
\label{sec:conclusions}

Slow mobile device hardware is a bottleneck to mobile Web performance.
We perform a large-scale measurement study to identify the impact of Android's battery-saver~(which throttles the CPU clock speed) mode on mobile Web performance.
Our data suggests that under low-battery conditions, sudden rises in page load time, total LongTask time, and time to interactive metrics are observed on some devices.
The average frame rate on some smartphones also declines, leading to unresponsive and paint-blocked websites.
\looseness -1




Through this paper we hope we motivate the need for new website design goals that improve mobile Web experience for slow mobile devices.
The Web performance community has developed numerous best practices for developers to deliver high-performance experiences to end users~\cite{grigorik2013high,souders1,souders2,amazonsilk,operamini,puffin,amp}. 
\looseness -1


\section*{Disclosure}
The positions, strategies, or opinions reflected in this article are those of the authors and do not necessarily represent the positions, strategies, or opinions of Akamai.

\vspace{-5pt}
\bibliographystyle{abbrv}
\bibliography{battery}
\end{document}